\documentclass[aps,pra,twocolumn,groupedaddress,showpacs,showkeys]{revtex4}

\usepackage{amssymb,amsmath,amsthm, pstricks}
\usepackage{graphicx}
\usepackage{color}

\newcommand{\tinyspace}{\mspace{1mu}}

\newcommand{\abs}[1]{\left\lvert\tinyspace #1 \tinyspace\right\rvert}

\def\squareforqed{\hbox{\rlap{$\sqcap$}$\sqcup$}}
\def\qed{\ifmmode\squareforqed\else{\unskip\nobreak\hfil
\penalty50\hskip1em\null\nobreak\hfil\squareforqed
\parfillskip=0pt\finalhyphendemerits=0\endgraf}\fi}

\newtheorem{theorem}{Theorem}
\newtheorem{lemma}[theorem]{Lemma}

\newcommand{\zo}{\{0, 1\}}

\newcommand{\eps}{\epsilon}

\def\squareforqed{\hbox{\rlap{$\sqcap$}$\sqcup$}}
\def\qed{\ifmmode\squareforqed\else{\unskip\nobreak\hfil
\penalty50\hskip1em\null\nobreak\hfil\squareforqed
\parfillskip=0pt\finalhyphendemerits=0\endgraf}\fi}

\begin{document}
\title{Optimal Protocols for Nonlocality Distillation}

\author{Peter H{\o}yer} \author{Jibran Rashid} 

\affiliation{Department of Computer Science, University of Calgary,
  2500 University Drive N.W., Calgary, Alberta, Canada T2N 1N4.}

\date{September 04, 2010}

\begin{abstract}
  Forster, Winkler, and Wolf recently showed that weak nonlocality can
  be amplified by giving the first protocol that distills a class of
  nonlocal boxes (NLBs) [Phys. Rev. Lett. {\bf{102}}, 120401 (2009)].
  We first show that their protocol is optimal among all non-adaptive
  protocols.  We next consider adaptive protocols.  We show that the
  depth $2$ protocol of Allcock \emph{et al.}  [Phys. Rev. A {\bf 80},
  062107, (2009)] performs better than previously known adaptive depth
  $2$ protocols for all symmetric NLBs.  We present a new depth $3$
  protocol that extends the known region of distillable NLBs.  We give
  examples of NLBs for which each of Forster \emph{et al.}'s, Allcock
  \emph{et al.}'s, and our protocol performs best.  The new
  understanding we develop is that there is no single optimal protocol
  for NLB distillation. The choice of which protocol to use depends on
  the noise parameters for the NLB.
\end{abstract}



\maketitle

Popescu and Rohrlich \cite{Popescu94b} proposed the hypothetical
nonlocal box (NLB) that attains the maximum value for the CHSH
inequality \cite{CHSH69} without allowing for communication between
spatially separated Alice and Bob.  It is natural to ask whether a
theory of reality can be maximally nonlocal.  Should we expect that
there exists another physical theory that allows for such
correlations? To understand why certain correlations are not allowed
by quantum physics it is necessary to understand the implications of
having such a correlation source.

Wim van Dam showed that a perfect nonlocal box implies trivial
communication complexity for boolean functions, i.e.\ any boolean
function may be computed by a single bit of communication between
Alice and Bob \cite{Dam00}.  This was extended by Brassard \textit{et
  al.} to include nonlocal boxes that work correctly with probability
greater than $\frac{3+\sqrt{6}}{6} \approx 0.908$ \cite{Brassard05}.

It was recently shown by Paw{\l}owski \textit{et al.}
\cite{Pawlowski09} that all strategies that violate Tsirelson's bound
\cite{Cirel80} also violate the principle of information causality
which states that the transmission of $n$ classical bits can cause an
information gain of at most $n$ bits.  It is not known whether this is
also true for nonlocal strategies that are prohibited by quantum
physics but do not violate Tsirelson's bound.

Is it possible to show that these results hold for all non-quantum
correlations? A positive answer would imply that quantum mechanics
restricts correlation sources that result in a world in which
surprisingly powerful information processing procedures would be
possible.  One attempt to solve this question is via nonlocality
distillation protocols.  The idea is to consider whether it is
possible for the players to concentrate the nonlocality in $n$ copies
of an imperfect nonlocal source to form a stronger nonlocal
correlation source.  In this sense it may be considered similar to
entanglement distillation.

Many of the known entanglement distillation protocols cannot be
utilized for nonlocality distillation since the former are allowed to
utilize both local operations and classical communication (LOCC)
whereas the latter are restricted to only local operations without any
communication.  Discussion of this approach and related results can be
found in \cite{SWolf08a,SWolf08b, SWolf09, Brunner09, Allcock09a,
  Allcock09b}.

Compared to entanglement distillation, nonlocality distillation
protocols are a recent development.  The first protocol for distilling
nonlocality was found by Forster, Winkler, and Wolf~\cite{SWolf09}.
They gave a non-adaptive protocol, which we define as a protocol in
which each NLB takes as input the original input to Alice and Bob, and
they derived an expression for the maximum value their distillation
protocol can achieve.  As our first result, we show that their
protocol is optimal among all non-adaptive protocols by proving a
matching lower bound.

Brunner and Skrzypczyk~\cite{Brunner09} next gave the first depth $2$
adaptive protocol which distills to a larger value than Forster
\textit{et al.}'s protocol for some NLBs.  Their protocol can be used
to distill to the asymptotic optimal value of~4 for NLBs that err in
exactly one of the four input cases, a class of NLBs which
\cite{Brunner09} coins \emph{correlated} NLBs.

This was next followed by Allcock \emph{et al.}~\cite{Allcock09b} who
gave an alternative depth $2$ adaptive protocol.  We show here that
the Allcock \emph{et al.} protocol distills the class of two parameter
(symmetric) NLBs considered in \cite{Brunner09} to a value strictly
bigger than the protocol in~\cite{Brunner09} attains, except in the
case of correlated NLBs for which both protocols distill to the
optimal value~4.

We then present a novel depth $3$ protocol that performs even better
for some NLBs.  Our protocol distills some NLBs that were not
previously known to be distillable, and it thus extends the known
region of distillable NLBs.

We finally show that for some NLBs, Forster \textit{et al.}'s original
protocol sometimes can distill to a value larger than both Allcock
\textit{et al.}'s and our protocols.  The picture that emerges is that
there is no known single optimal protocol for NLB distillation.  Which
protocol to apply depends on the parameters of the given NLB.  We
conclude that our understanding of nonlocality distillation is still
in its infancy and there is still plenty to be discovered about
nonlocal boxes.

\begin{figure}[t]
\includegraphics[scale=0.6]{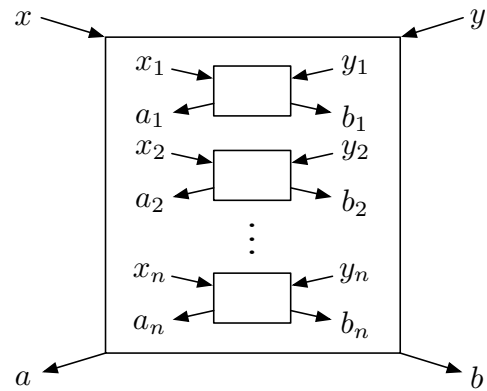}   
\caption[A depth $n$ NLB distillation protocol.]{NLB distillation protocol 
   of depth $n$}
\label{figure:generic}
\end{figure}

\textit{Framework.} Consider two spatially separated parties Alice and
Bob who receive input bits $x$ and $y$ from a uniform distribution.
For the CHSH inequality, the players are required to produce output
bits $a$ and $b$, respectively, such that $a \oplus b = xy$.  The
matrix ${\bf p}$, with its rows indexed by $xy$ and columns by $ab$,
gives the probability with which Alice and Bob output $a$ and $b$ on
inputs $x$ and $y$, respectively.  In addition to positivity and
normalization, the no-signalling conditions are enforced on ${\bf p}$,
so the local marginal distribution of Alice is independent of the
output of Bob and vice versa.  The value that the CHSH inequality
takes for a strategy ${\bf p}$ is given by
\begin{eqnarray}
\label{intro:chsheq}
V({\bf p}) = \sum_{a\oplus b = xy}  p_{ab|xy} - \sum_{a\oplus b \neq xy}  p_{ab|xy}.
\end{eqnarray}
The perfect nonlocal box is defined to output a uniform distribution
over the bits $a$ and $b$ on inputs $x$ and $y$ such that $a \oplus b
= xy.$ We consider the following general NLB as a correlation resource
for the CHSH inequality
\begin{displaymath}
{\bf p}=
\frac{1}{4} 
\left( \begin{array}{c@{+}cc@{-}cc@{-}cc@{+}c}
1 & \delta_1 & 1 & \delta_1 & 1 & \delta_1 & 1 & \delta_1 \\ 
1 & \delta_2 & 1 & \delta_2 & 1 & \delta_2 & 1 & \delta_2 \\ 
1 & \delta_3 & 1 & \delta_3 & 1 & \delta_3 & 1 & \delta_3 \\
1 & \eps     & 1 & \eps     & 1 & \eps     & 1 & \eps \\
\end{array} \right),
\end{displaymath}
where the parameters $\delta_1, \delta_3, \delta_3,$ and $\eps$ are in
$\left[-1,1\right]$.  To remove redundancy and focus on the key terms
in the distribution, we shall write the NLB as
\begin{equation*}
\frac{1}{4} 
(1 + \delta_1,
1 + \delta_2,
1 + \delta_3,
1 + \epsilon)^T.
\end{equation*}
A single usage of the NLB gives us a value of 
\begin{displaymath}
V({\bf p}) = \delta_1+\delta_2+\delta_3 - \epsilon.
\end{displaymath}
We are interested in distilling this NLB resource $\bf p$ such that
the distilled NLB attains a greater value than the original value
$V({\bf p})$.  A~distillation protocol takes as input the original two
input bits $x$, $y$ of Alice and Bob and $n$ identical copies of a NLB
${\bf p}$, and it outputs two bits $a$ and~$b$.  See
Figure~\ref{figure:generic}.  The protocol specifies what each of
Alice and Bob input to each of the $n$ NLBs.  Alice's input $x_1$ to
the first NLB can depend only on her original input bit~$x$.  Her
input $x_2$ to the second NLB can depend on her original input bit~$x$
and her output $a_1$ of the first NLB, and so forth.  After receiving
all her $n$ output bits $a_1, a_2, \ldots, a_n$, Alice then outputs a
bit $a$ that can depend on $x$ and $a_1, a_2, \ldots, a_n$.
Similarly, Bob's input $y_1$ to the first NLB can depend only on his
original input bit~$y$.  His input $y_2$ to the second NLB can depend
on his original input bit~$y$ and his output $b_1$ of the first NLB,
and so forth.  He also outputs a bit $b$ which may depend on $y$ and
$b_1, b_2, \ldots, b_n$.  The goal is for Alice and Bob to have that
$a \oplus b = x y$.

We assume, as is common in communication complexity, that both players
know the four parameters $\delta_1, \delta_2, \delta_3, \epsilon$ that
specify their NLB.  We also assume that Alice and Bob give their $n$
input bits to the $n$ NLBs in the same order.  This is strictly not
necessary for the model to be well-defined and to be of interest.  In
this paper, however, we do not consider that more general model.

We refer to $n$, the number of NLBs, as the \emph{depth} of the
protocol.  We note that for any depth $n$ protocol, there is a depth
$n+1$ protocol that achieves the same value: this can for instance be
obtained by Alice and Bob each inputting an arbitrary bit to NLB
number $n+1$ and disregarding the output bits $a_{n+1}$ and $b_{n+1}$.
When we thus talk about the class of depth $n$ protocols, this
includes protocols equivalent to all protocols of depth less than~$n$.
Conversely, for \emph{some} depth $n$ protocols, there exists a depth
$n-1$ achieving the same value.

The goal of NLB distillation is given $n$ identical NLBs to obtain a
NLB that achieves as high a value as possible.

\begin{figure}[t]
\includegraphics[scale=0.6]{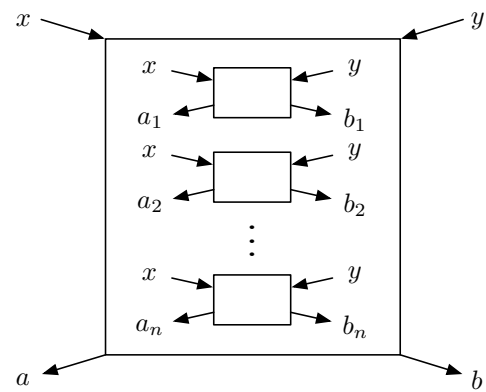}   
\caption[Non-adaptive protocol.]{Non-adaptive protocol.   When $a=a_1
  \oplus a_2 \oplus \cdots \oplus a_n$, and $b=b_1 \oplus b_2 \oplus
  \cdots \oplus b_n$, we refer to the protocol as Forster \emph{et
    al.}'s parity protocol.}
\label{figure:non-adaptive}
\end{figure}

We say a protocol is \emph{non-adaptive} if each of the $n$ NLBs takes
as input the original input bits $x$ and $y$ received by Alice and
Bob.  See Figure~\ref{figure:non-adaptive}.  A~non-adaptive protocol
allows for parallelism and can be implemented by each of Alice and Bob
inputting their $n$ inputs to the $n$ NLBs in parallel.
A~non-adaptive protocol can be implemented in a single round.  In
contrast, in an \emph{adaptive} distillation protocol, the two players
may choose to use the output of previous NLBs to determine the input
to later NLBs.  Restricting the number of rounds is well-studied in
communication complexity and other related settings.  In classical
communication complexity, the seminal paper~\cite{KW} connects bounded
round protocols to circuits to prove a lower bound on the circuit
complexity of the computational problem of graph connectivity.  In
quantum communication complexity, two early results on protocols of
bounded rounds are~\cite{KNTZ01, JRS03}.


\textit{Parity is Optimal.} Forster \textit{et al.}~\cite{SWolf09}
derived an expression for the maximum value their distillation
protocol can achieve.  We show that the parity protocol (Figure
\ref{figure:non-adaptive}) considered by Forster \textit{et
  al.}~\cite{SWolf09} is an optimal non-adaptive distillation
protocol.  We show this by first determining the expression for the
value attained by the parity protocol over $n$ NLBs.  We then show
that the value attainable by any non-adaptive protocol over $n$ NLBs
is never greater than the the value attained by the parity protocol
over $k$ NLBs such that $1 \leq k \leq n$.
\begin{theorem}
\label{thm1}
Forster \textit{et al.}'s parity protocol~\cite{SWolf09} is optimal
among all non-adaptive distillation protocols.
\end{theorem}
The upper bound is a simple generalization of Theorem~2
in~\cite{SWolf09} which can be proved by considering the parity of the
number of heads obtained by flipping a coin with bias $\delta$ a
number of $k$ times.
\begin{lemma}[\cite{SWolf09}]
\label{lm1}
The parity protocol over $n$ NLBs attains the value of
$\delta_1^n+\delta_2^n+\delta_3^n-\eps^n$ for the CHSH inequality.
\end{lemma}
We now give a matching lower bound by showing that the value attained
by any other non-adaptive distillation protocol is upper bounded by
the value obtained the parity protocol over a chosen number of NLBs.
\begin{lemma}
\label{lm2}
The value attainable by any non-adaptive protocol using at most $n$
NLBs is upper bounded by $\max_{1 \leq k \leq n}
\abs{\delta_1^k+\delta_2^k+\delta_3^k-\eps^k}$.
\end{lemma}

\begin{proof}[Proof of Lemma \ref{lm2}]
  Let the $n$ bit pairs $\left(a_{i},b_{i}\right)$ be the output of
  the $n$ NLBs that Alice and Bob obtain for inputs $x$ and $y$,
  respectively.  See Figure~\ref{figure:non-adaptive}.  The pair
  $\left( a_{i},b_{i}\right)$ is drawn from $\{00,01,10,11\}$ with
  respect to the distribution $\mu = \frac{1}{4} \{
  1+\delta,1-\delta,1-\delta,1+\delta \}$, where $\delta$ is the bias
  for the row corresponding to the inputs received by the players.
  For inputs $x$ and $y$, let $A, B \subseteq \zo^n$ be the set of
  strings for which Alice and Bob's final output is~$1$, respectively.

  Given that Alice and Bob input bits $x$ and $y$ into the $n$ NLBs,
  the probability that they receive bit strings $a$ and $b$ of length
  $n$, respectively, from the NLBs is given by
\begin{align*}
p_{ab|xy} & =  \prod_{i=1}^{n} \left( \frac{1-\delta}{4} + \frac{\delta}{2} \left[a_{i}=b_{i} \right] \right) \\
 & = \left( \frac{1-\delta}{4} \right)^{n} \prod_{i=1}^{n} \left( 1 + \frac{2\delta}{1-\delta} \left[a_{i}=b_{i} \right] \right) \\
  & =  \left( \frac{1-\delta}{4} \right)^{n} \left( \frac{1+\delta}{1-\delta}\right)^{n-\abs{a \oplus b}} \\ 
  & =  \frac{1}{4^n} \left( 1-\delta\right)^{\abs{a \oplus b}}  \left( 1+\delta \right)^{n-\abs{a \oplus b}},
\end{align*}
where $\left[a_{i}=b_{i} \right]$ = $1$ if $a_i=b_i$ and $0$
otherwise.  The probability of obtaining output $11$ is given by
\begin{align*}
& q_{\left(A,B\right)} \left(\delta \right) \\
&= \frac{1}{4^n} \sum_{a \in A} \sum_{b \in B} (1-\delta)^{|a \oplus b|}
   (1+\delta)^{n-|a \oplus b|} \\
&= \frac{1}{4^n} \sum_{a \in A} \sum_{b \in B} \sum_{z \in \zo^n} 
   \chi_z(a \oplus b) \delta^{|z|} \\
&= \frac{1}{4^n} \sum_{z \in \zo^n} \delta^{|z|} 
   \sum_{a \in A} \sum_{b \in B} \chi_z(a \oplus b) \\
&= \sum_{z \in \zo^n} \delta^{|z|} 
   \left( \sum_{a \in A} \frac{1}{2^n} \chi_z(a) \right) 
   \left( \sum_{b \in B} \frac{1}{2^n} \chi_z(b) \right)  \\
&= \sum_{z \in \zo^n} \delta^{|z|} 
   \left( \sum_{s} \frac{1}{2^n} \chi_z(s) \left[s \in A \right] \right)\\ 
& \hphantom{=} \times \left( \sum_{t} \frac{1}{2^n} \chi_z(t) \left[t \in B \right] \right)\\  
&= \sum_{z \in \zo^n} \delta^{|z|} 
   \left( \sum_{s} \frac{1}{2^n} \chi_z(s) \left(\frac{f(s)+1}{2}\right) \right)\\ 
& \hphantom{=} \times \left( \sum_{t} \frac{1}{2^n} \chi_z(t) \left(\frac{g(t)+1}{2}\right) \right)\\
&= \sum_{z \in \zo^n} \delta^{|z|} 
   \left(\frac{\hat{f}_z + \left[z=0\right]}{2}\right) \left(\frac{\hat{g}_z + \left[z=0\right]}{2}\right) \\
&= \sum_{z \in \zo^n} \frac{\delta^{|z|}}{4} 
   \left(\hat{f}_z\hat{g}_z + \left( 1+\hat{f}_0+\hat{g}_0 \right)\left[z=0\right]\right).
\end{align*}
Here $\chi_z(a \oplus b) = (-1)^{z\cdot(a\oplus b)}$ is a character
for the group $\mathbb{Z}_{2^n}$ and $f$ and $g$ are $+1$ when $s$ and
$t$ are in $A$ and $B$, respectively, and $-1$ otherwise.  To see that
the second equation follows from the first, expand the inner-most
product $(1-\delta)^{|a \oplus b|} (1+\delta)^{n-|a \oplus b|}$ in its
$2^n$ terms and then rewrite each of those as the evaluation of $a
\oplus b$ on one of the $2^n$ characters~$\chi_z$.  For the second
last equality, notice that the sum $\sum_s \frac{1}{2^n} \chi_z(s)
f(s)$ equals the Fourier coefficient~$\hat f_z$.

The probability of obtaining output $00$ is the same as the expression
for $11$, except that the sign in front of each of $\hat{f}_0$ and
$\hat{g}_0$ gets flipped.  Then, the probability of obtaining output
$00$ or $11$ is given by
\begin{align*}
r_{\left(A,B\right)} \left(\delta \right)=\frac{1}{2} \left(1 + \sum_{z \in \zo^n} \hat{f}_z\hat{g}_z \delta^{|z|} \right).
\end{align*}
We use this expression to determine a bound on the value $V({\bf p})$
that any non-adaptive distillation protocol may attain for a NLB,
given the biases $\delta_1, \delta_2, \delta_3$, and $\eps$.
\begin{align*}
V({\bf p})  
& = \Big( \sum_i \left( 2r(\delta_i)-1 \right)\Big)- \left( 2r(\epsilon)-1 \right)  \\
& = \sum_{z \in \zo^n} \hat{f}_z\hat{g}_z \left( {\delta_1}^{|z|}+{\delta_2}^{|z|}+{\delta_3}^{|z|} - \epsilon^{|z|} \right) \\
& \leq \sum_{z \in \zo^n} \abs{\hat{f}_z} \cdot \abs{\hat{g}_z} \cdot \abs{ {\delta_1}^{|z|}+{\delta_2}^{|z|}+{\delta_3}^{|z|} - \epsilon^{|z|}} \\
& \leq \max_{k} \abs{ {\delta_1}^k+{\delta_2}^k+{\delta_3}^k - \epsilon^k} \sum_{z \in \zo^n} \abs{\hat{f}_z} \cdot \abs{\hat{g}_z} \\
& \leq \max_{k} \abs{ {\delta_1}^k+{\delta_2}^k+{\delta_3}^k - \epsilon^k},
\end{align*}
where the last inequality follows from $\hat{f}_z$ and $\hat{g}_z$
being normalized functions.
\end{proof}
We conclude that Forster \textit{et al.}'s parity protocol is an
optimal non-adaptive protocol.  We note that Alice and Bob perform
identical operations in the parity protocol.  In contrast, when
allowing for adaptive protocols, an optimal protocol does not
necessarily imply that Alice and Bob perform identical operations.

\begin{figure}[t]
\includegraphics[scale=0.6]{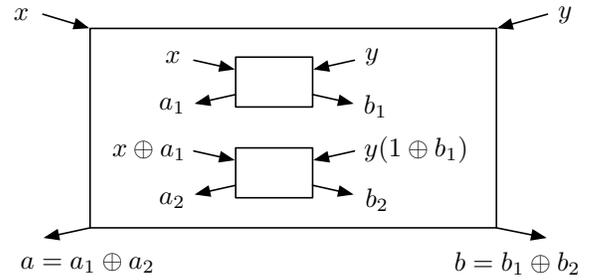}
\caption[Adaptive protocol.]{The depth $2$ adaptive distillation
  protocol of Allcock \textit{et al.}~\cite{Allcock09b}.}
\label{figure:2query} 
\end{figure}

\begin{figure}[t] \vspace{\baselineskip}
\includegraphics[scale=.31]{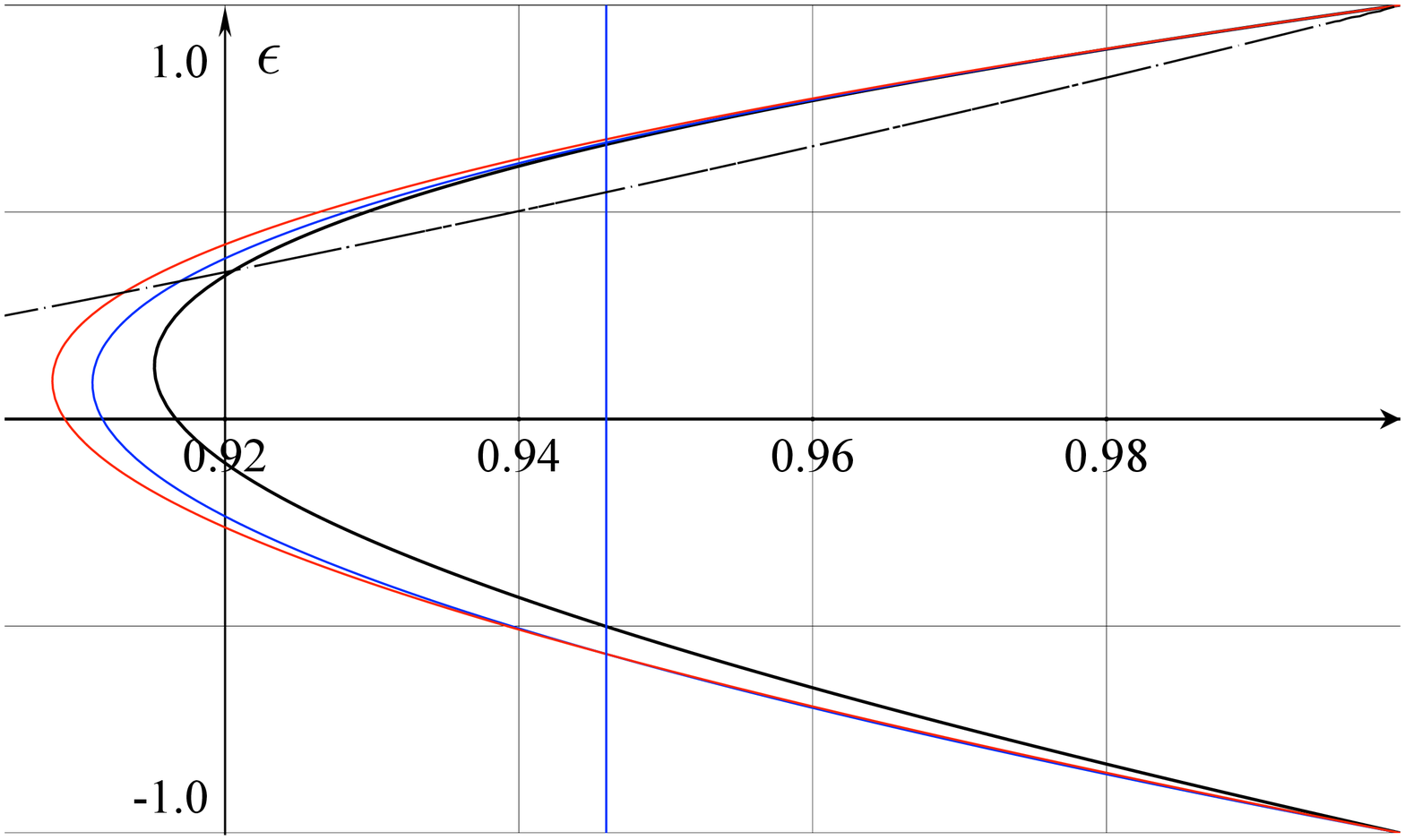}  
\caption{NLBs distilled by the protocols.  The outermost curve (red in
  the online version) represents Allcock \emph{et al.}'s depth $2$
  adaptive protocol.  The curve in the middle (blue in the online
  version) is our new depth $3$ protocol, and the innermost curve
  (black in the online version) is the depth $2$ adaptive parity
  protocol.  The region above the black dotted curve that goes through
  the point $\delta=\cos(\pi/9) \approx 0.94$ and
  $\epsilon=\frac{1}{2}$ represents distributions that are obtainable
  within quantum theory.  The vertical line is $\delta=35/37$.  The
  two outermost curves cross at the point $\delta=35/37$ and $\epsilon
  = -21/37$.}
\label{fig:distillcurve}
\end{figure}

\textit{Adaptive Distillation Protocols.} Brunner and Skrzypczyk
\cite{Brunner09} consider an adaptive distillation protocol of depth
two that asymptotically distills correlated NLBs to the maximum value
of $4$.  We refer to their protocol as the adaptive parity protocol.
The class of \emph{correlated NLBs} is given by
\begin{equation}
 \frac{1}{4} 
\left( \begin{array}{cccc}
2 & 0 & 0 & 2 \\ 
2 & 0 & 0 & 2 \\ 
2 & 0 & 0 & 2 \\
1+\epsilon & 1-\epsilon & 1-\epsilon & 1+\epsilon \\
\end{array} \right),
\end{equation}
where $\epsilon \in \left[-1,1\right]$ and $\delta_1 =
\delta_2=\delta_3 =1$.  The value attained by this protocol is
$\frac{1}{4}\left(13-4\eps-\eps^2 \right)$.  We briefly present a
depth $k$ version of the above protocol, which also illustrates the
intuition behind why it works.  The players input their bits $x$ and
$y$ to the first NLB.  The input to the $i^{\textrm{th}}$ NLB, for $i
> 1$, is given by the logical \textsf{AND} of the original input bit
and the parity of the $i-1$ output bits obtained from the previous
NLBs.  The final output for a depth $k$ protocol is the parity of
their $k$ output bits received from the NLBs.  Let $p =
\frac{1+\epsilon}{4}$.
\begin{theorem} 
\label{thm2}
The depth $k$ adaptive parity protocol attains the value
$4\left(1-p\left(p+\frac{1}{2} \right)^{k-1} \right)$ on correlated
NLBs.
\end{theorem}
\begin{proof}
  For inputs $xy \in \{ 00, 01, 10 \}$ we always obtain output bits
  $a$ and $b$ with even parity.  For the case when $xy$ is equal to
  $11$, consider the first NLB which outputs bits with odd parity.
  Let this be the $i^{\textrm{th}}$ NLB with output bits $a_i$ and
  $b_i$ for $i \geq 1$.  This implies that the input to the
  ${(i+1)}^{\textrm{th}}$ NLB has odd parity and this guarantees that
  all NLBs at depth greater than $i$ output even parity.  Therefore,
  the final output parity will be odd due to $a_i$ and $b_i$.  This
  implies that for $-1 \leq \epsilon < 1$, the protocol
  asymp\-to\-ti\-cally distills all the corresponding NLBs arbitrarily
  close to a perfect NLB.  For $k=2$ we obtain
\begin{displaymath}
p_{00|11} = p\left( p + \frac{1}{2} \right), 
\end{displaymath} 
which implies a ratio of $p + \frac{1}{2}$ between the probability of
the distilled and original NLBs.  For a depth $k$ protocol, the
probability to obtain odd parity output, given $xy = 11$, is
$1-2p\left(p+\frac{1}{2}\right)^{k-1}$.  This leads to a distilled NLB
that attains the required value.
\end{proof}
Here we consider the more general class of \emph{symmetric NLBs} given
by $\delta_1 = \delta_2=\delta_3$, which we represent by
\begin{eqnarray}
\label{symnlb}
\frac{1}{4} 
\left( \begin{array}{c}
1+\delta   \\
1+\epsilon \\
\end{array} \right).
\end{eqnarray}
These NLBs correspond to the two parameter family of states considered
by Brunner and Skrzypczyk \cite{Brunner09}.  All correlated NLBs are
symmetric, but not vice-versa.  To specify a depth 2 protocol for
symmetric NLBs, we only need to provide the mapping for
(\ref{symnlb}).  Brunner and Skrzypczyk's protocol gives the following
mapping
\begin{equation}
\frac{1}{4} 
\left( \begin{array}{c}
1+\delta \\ 
1+\epsilon	
\end{array} \right) \mapsto
\frac{1}{4} 
\left( \begin{array}{c}
1+\delta^2   \\
\frac{\epsilon^2+\epsilon+3\epsilon \delta-\delta+4}{4}
\end{array} \right).
\label{eq:bs-entries}
\end{equation}
The value attained is $\frac{1}{4}\left(12\delta^2+\delta
  -3\eps\delta-\eps-\eps^2 \right)$.  Allcock \textit{et
  al.}~\cite{Allcock09b} next gave a protocol that we now show
performs better than the above protocol for the entire class of
symmetric NLBs.  We use the representation in
Figure~\ref{figure:2query} of their protocol.
The mapping for Allcock \textit{et al.}'s protocol is
\begin{equation}
\frac{1}{4} 
\left( \begin{array}{c}
1+\delta \\ 
1+\epsilon	
\end{array} \right) \mapsto
\frac{1}{4} 
\left( \begin{array}{c}
1+\delta^2   \\
\frac{3\delta^2+\delta+\epsilon \delta-\epsilon+4}{4}\\
1+\delta^2   \\
\frac{\epsilon^2+\epsilon+3\epsilon \delta-\delta+4}{4}
\end{array} \right).
\label{eq:our-entries}
\end{equation}
The value attained by this protocol is
$\frac{1}{4}\left(11\delta^2+2\delta -2\eps\delta-2\eps-\eps^2
\right)$.  The first, third, and forth entries on the right hand side
in Eq.~\ref{eq:our-entries} are identical to the corresponding entries
obtained by Brunner and Skrzypczyk's protocol given in
Eq.~\ref{eq:bs-entries}.  The second entry is strictly greater than
$\frac{1+\delta^2}{4}$ whenever $\epsilon < \delta < 1$.  If
$\delta=1$, the second entry is the same as in
Eq.~\ref{eq:bs-entries}.  If $\delta \leq \epsilon$, the output
distribution of the NLB can be simulated by quantum mechanics and does
thus not represent nonlocality (see \cite{Navascues08,Cirel80}).  We
conclude that the Allcock \textit{et al.} protocol is strictly better
than Brunner and Skrzypczyk's protocol for all symmetric
non-correlated NLBs.  Further, Allcock \textit{et al.}'s protocol
distills some NLBs that are not distillable by Forster \textit{et
  al.}'s and Brunner and Skrzypczyk's protocols as shown in
Figure~\ref{fig:distillcurve}.
 
\textit{New depth $3$ protocol.} Similar to the non-adaptive case, we
may ask whether Allcock \textit{et al.}'s protocol is an optimal
adaptive protocol for general NLBs.  Since that protocol maps out of
the class of symmetric NLBs, we cannot use the above arguments to show
optimality for general NLBs and arbitrary depth protocols.  We also
find that a local permutation of the protocol performs better for
certain non-symmetric NLBs~\footnote{The protocol introduced in
  Section III of Allcock \emph{et al.}'s paper~\cite{Allcock09b} is
  given in this form.}.  The inputs to the second NLB in this protocol
are given by 
\begin{equation}
\label{lvar}
f_2 = x a_1 \text{ and } g_2 = 1\oplus y \oplus b_1.
\end{equation}
The permutation does not simply interchange the roles of Alice and Bob
and is dependent on the biases $\delta_2$ and $\delta_3$ as shown in
Table~\ref{table:nlbsim}.  Numerical simulations indicate that one of
these two protocols always performs better or as well as the adaptive
parity protocol for the entire class of non-quantum NLBs.
Table~\ref{table:nlbsim} presents different choices of NLB parameters,
such that no single depth 2 NLB protocol is always optimal.  There
even exist situations for which the non-adaptive parity protocol
performs better than the depth $2$ adaptive protocols.

\begin{table}
 \centering
 \begin{tabular}{|c|c|c|r @{.} l|r @{.} l|r @{.} l|r @{.} l|r @{.} l|r @{.} l|} \hline
$\delta_1$ &  $\delta_2$ &  $\delta_3$& \multicolumn{2}{|c|}{$\epsilon$}& \multicolumn{2}{|c|}{$P$} &\multicolumn{2}{|c|}{$P_{\oplus}$} & \multicolumn{2}{|c|}{$P_{\text{BS}}$}& \multicolumn{2}{|c|}{$P_{\text{A}}$}& \multicolumn{2}{|c|}{$P_{\text{perm}}$}  \\   \hline
$0.92$ & $0.92$ & $0.92$ &$-0$&$22$ & $2$&$98$ &$2$&$4908$& $2$&$9639$ & $\bf{2}$&$\bf{9867}$ & $\bf{2}$&$\bf{9867}$  \\
$0.96$ & $0.84$ & $0.96$ &$0$&$24$ & $2$&$52$ &$2$&$4912$& $2$&$5692$ & $\bf{2}$&$\bf{5932}$ & $2$&$4600$ \\
$0.96$ & $0.96$ & $0.84$ &$0$&$24$ & $2$&$52$ &$2$&$4912$& $2$&$5692$ & $2$&$4600$ & $\bf{2}$&$\bf{5932}$ \\
$0.96$ & $0.96$ & $0.96$ &$0$&$60$ & $2$&$28$ &$\bf{2}$&$\bf{4048}$& $2$&$3328$ & $2$&$3364$& $2$&$3364$  \\
\hline
\end{tabular}
\caption{\textnormal{
    Values for NLB distillation protocols of depth $2$.   Column $P$ is the
    value of the nonlocal box itself, $P_{\oplus}$ is non-adaptive parity,
    $P_{\text{BS}}$ is adaptive parity, $P_{\text{A}}$ is the protocol of
    Allcock \textit{et al.} and $P_{\text{perm}}$ is its local variant
    given in Eq.~\ref{lvar}.
  }}
\label{table:nlbsim}
\end{table}

\begin{table}
  \centering
  \begin{tabular}{|c|r @{.} l|r @{.} l|r @{.} l|r @{.} l|r @{.} l|r @{.} l|r @{.} l|r @{.} l|r @{.} l|r @{.} l|}
   \hline
$\delta$                    &  \multicolumn{2}{|c|}{$\epsilon$}& \multicolumn{2}{|c|}{$P$}   & \multicolumn{2}{|c|}{$P_{\oplus}$} & \multicolumn{2}{|c|}{$P_{A}$} & \multicolumn{2}{|c|}{$P_{3}$}&  \multicolumn{2}{|c|}{$P_{6}$}    & \multicolumn{2}{|c|}{$P_{\text{new}}$}  \\
   \hline
$0.96$ & $-0$&$48$ & $3$&$36$& $3$&$3600$& $3$&$4272$ & $3$&$4399$ & $3$&$3375$ &  $\bf{3}$&$\bf{4907}$ \\
$0.96$ & $0$&$60$ & $2$&$28$& $2$&$4382$& $2$&$3364$ & $2$&$3786$ &  $\bf{2}$&$\bf{4394}$ & $2$&$3864$ \\
$0.92$ & $-0$&$22$ & $2$&$98$& $2$&$4908$& $\bf{2}$&$\bf{9867}$ & $2$&$9490$ &  $2$&$7308$ & $2$&$9842$ \\
\hline
\end{tabular}
\caption{\textnormal{
    Values for NLB distillation protocols of various depths.   Column $P$
    is the value of the nonlocal box itself, $P_{\oplus}$ is optimal-depth
    non-adaptive parity, $P_{\text{BS}}$ is adaptive parity,
    $P_{\text{A}}$ is the protocol of Allcock \textit{et al.}, $P_{3},
    P_{6}$ are our generalizations thereof to depths $3$ and $6$,
    respectively, and $P_{\text{new}}$ is the protocol given by
    Eq.~\ref{prot3} below.}}
\label{table:nlbsim1}
\end{table}

We may consider a generalization of the Allcock \textit{et al.}'s
protocol to arbitrary depth $n$, where the input to the
$k^{\text{th}}$ NLB, for $k>1$ is given by
\begin{equation}
f_k = x \oplus \bigoplus_{i=1}^{k-1} a_i \ \text{ and }\ g_k = y\Big(1 \oplus \bigoplus_{i=1}^{k-1} b_i\Big).
\end{equation}
We find that this does not yield an optimal protocol, since for depth
$3$ a better protocol exists, that uses the same inputs to the first
two NLBs as in Figure~\ref{figure:2query} and with inputs to the third
NLB given by
\begin{equation}
\label{prot3}
\begin{array}{cl}
f_3 &= a_2(a_1\oplus 1) \oplus x(a_1\oplus a_2 \oplus a_1a_2) \\
g_3 &= 1 \oplus b_1 \oplus b_2(1 \oplus b_1) \oplus y(1 \oplus b_2 \oplus b_1b_2).
\end{array}
\end{equation}
This protocol attains the following value for symmetric NLBs
\begin{equation}
\label{value3}
\frac{1}{16}\left( 39\delta^3 + \delta^2 \left( \epsilon+16  \right)+ \delta\left(1-16 \epsilon-8\epsilon^2 \right)-\epsilon\right).
\end{equation}
Figure~\ref{fig:distillcurve} shows the regions distilled by the above
known protocols.  The region above the black dotted curve represents
NLBs with output distributions that are simulatable within quantum
theory~\cite{Cirel80}.  The three convex sets, each bounded by one of
the three similar curves represents the NLBs that are distilled by
each of the three protocols.  The outermost curve (red in the online
version) is Allcock \emph{et al.}'s depth $2$ protocol.  The curve in
the middle (blue in the online version) represents our depth $3$
protocol, and the innermost curve (black in the online version)
represents adaptive parity of depth~$2$.

Interestingly, the lower boundary of $\delta$ for which Allcock
\emph{et al.}'s protocol distills symmetric NLBs is exactly
$\frac{3+\sqrt{6}}{6}$.  Further, $\frac{3+\sqrt{6}}{6}$ is also the
lower boundary value of $\delta$ for which Forster's protocol distills
symmetric NLBs.  No known protocol distills symmetric NLBs with
$\delta \leq \frac{3+\sqrt{6}}{6}$.

Our new depth $3$ protocol extends the known region of distillable
NLBs.  When $\delta > \frac{35}{37}$, our new protocol distills for a
value of $\epsilon$ strictly smaller than what is distillable by
Allcock \emph{et al.}'s protocol.  When e.g.{} $\delta=0.95$ and
$\epsilon=-0.607$, our protocol distills as the only protocol among
the protocols discussed in this paper.  Further, our new protocol
attains a higher value for some NLBs within its distillable region.
The values in Table \ref{table:nlbsim1} again reinforce the notion
that there is no single optimal NLB distillation protocol.
  
\begin{acknowledgments}
  Discussions with Alain Tapp, C\u{a}t\u{a}lin Dohotaru, Nathan Wiebe
  and Donny Cheung.  We also thank the anonymous referees for
  insightful and thoughtful comments.  This work was supported by
  Canada's Natural Sciences and Engineering Research Council (NSERC),
  the Canadian Network Centres of Excellence for Mathematics of
  Information Technology and Complex Systems (MITACS), and
  QuantumWorks.  P.H.{} is a Scholar of the Canadian Institute for
  Advanced Research (CIFAR).
\end{acknowledgments}


\begin{thebibliography}{99}

\bibitem{Popescu94b}
S.~Popescu and D.~Rohrlich,
\newblock  Found. Phys. {\bf 24}, 379, (1994).

\bibitem{CHSH69}
J.~F. Clauser, M.~A. Horne, A.~Shimony, and R.~A. Holt,
\newblock Phys. Rev. Lett. {\bf 23}, 880, (1969).

\bibitem{Dam00}
W.~van Dam,
\newblock PhD thesis, University of Oxford, (2000).

\bibitem{Brassard05}
G.~Brassard, H.~Buhrman, N.~Linden, A.~A. M\'ethot, A.~Tapp, and F.~Unger,
\newblock Phys. Rev. Lett. {\bf 96}, 250401, (2006).

\bibitem{Pawlowski09}
M.~Paw{\l}owski, T.~Paterek, D.~Kaszlikowski, V.~Scarani, A.~Winter, and
  M.~{\.Z}ukowski,
\newblock Nature {\bf 461}, 1101, (2009).

\bibitem{Cirel80}
B.~S. Cirel’son,
\newblock  Lett. Math. Phys. {\bf 4}, 93, (1980).

\bibitem{SWolf08a}
D.~Dukaric and S.~Wolf,
\newblock arXiv:quant-ph/0808.3317v2.
  
\bibitem{SWolf08b}
M.~Fitzi, E.~H\"{a}nggi, V.~Scarani, and S.~Wolf,
Proceedings of Quantum Communication, Measurement 
and Computing (QCMC), (2008).

\bibitem{SWolf09}
M.~Forster, S.~Winkler, and S.~Wolf,
\newblock Phys. Rev. Lett. {\bf 102}, 120401, (2009).

\bibitem{Brunner09}
N.~Brunner and P.~Skrzypczyk,
\newblock Phys. Rev. Lett. {\bf 102}, 160403, (2009).

\bibitem{Allcock09a}
J.~Allcock, N.~Brunner, M.~Paw{\l}owski, and V.~Scarani,
\newblock Phys. Rev. A {\bf 80}, 040103(R), (2009).

\bibitem{Allcock09b}
J.~Allcock, N.~Brunner, N.~Linden, S.~Popescu, P.~Skrzypczyk, and T.~V\'ertesi,
\newblock Phys. Rev. A {\bf 80}, 062107, (2009).

\bibitem{KW}
M. Karchmer and A. Wigderson,
\newblock SIAM J. Discrete Math. {\bf 3}, 255--265 (1990).

\bibitem{KNTZ01}
H. Klauck, A. Nayak, A. Ta-Shma, and D. Zuckerman,
\newblock  STOC 2001: Proceedings of the Thirty-third 
  Annual ACM Symposium on Theory of Computing, 
  pp. 124--133 (2001).

\bibitem{JRS03}
R. Jain, J. Radhakrishnan, and P. Sen,
\newblock 
  FOCS 2003: Proceedings of the 
  Fourty-forth Annual IEEE Symposium on Foundations of Computer Science, 
  pp. 220--229 (2003).

\bibitem{Navascues08}
M.~Navascu\'es, S.~Pironio, and A.~Ac\'in,
\newblock New J. Phys. {\bf 10}, 073013, (2008).


\end{thebibliography}
\end{document}